# Melt Grown ZnO Bulk Crystals

Detlev Schulz, Steffen Ganschow, Detlef Klimm
Leibniz Institute for Crystal Growth, D – 12489 Berlin, Max-Born-Str. 2, GERMANY

## ABSTRACT

Bulk crystals of zinc oxide can be grown from the melt by a Bridgman technique under pressure. This new technology using an iridium crucible shows the potential to yield large single crystals of good crystalline perfection. Crystals with diameters up to 33 mm and a length of up to 50 mm have been demonstrated. The impurity content can be strongly reduced by using the crucibles repeatedly.

## INTRODUCTION

Commercial zinc oxide wafers are fabricated almost exclusively from hydrothermally grown ZnO bulk crystals. The hydrothermal technology rests on decades of experience mainly with the growth of α-quartz crystals for piezoelectric applications. From alkaline hydrothermal solutions of zinc oxide, ZnO crystals sized up to 3 inch can be grown within several weeks. It is typical for crystal growth processes from solutions, however, that traces of the solvent (here mainly H, Li, K) are incorporated to the grown bulk. This is uncritical for a piezoelectric material, but can have substantial influence on a semiconductor.

Crystal growth technologies that rely on pure melts without solvent (e.g. Czochralski, Bridgman) are desirable alternatives, and are used almost exclusively for the mass production of other electronic materials (Si, Ge, GaAs). For a long time it was assumed that the high melting point of ZnO ($T_f$ = 1975 °C), together with the high oxygen partial pressure that is needed to stabilize ZnO at $T_f$, would not allow to find any crucible material that can withstand molten ZnO. Recently we could show that this claim is wrong, instead a reactive atmosphere containing carbon dioxide can deliver by the equilibrium reaction $CO_2 \leftrightarrow CO + ½ O_2$ a "self adjusting" oxygen partial pressure which stabilizes ZnO melt in an iridium crucible [1]. This way zinc oxide bulk crystals can be grown from iridium crucibles e.g. by the Bridgman method.

Now boules with 33 mm diameter and ca. 50 mm length can be grown that are suitable for the production of wafers that were successfully used for the deposition of (Zn,Mg)O and (Zn,Cd)O epilayers by MBE [2]. As expected, the concentration of most impurities is lower, compared with hydrothermal samples [3].

Despite these encouraging results, two major problems arise:
a) Thermal stresses during crystallization and cooling of the ZnO boule might be so large that grain boundaries or even cracks will form.
b) Some trace impurities that are still found in the melt grown crystals (e.g. Fe), are presumably introduced from the crucible: The chemical purity of the iridium metal is typically only 99.9%.

This contribution compares the status of ZnO melt growth with other technologies, preferably hydrothermal growth, and discusses prospects of future improvements.

# EXPERIMENT

A Bridgman-like setup is used for the growth of zinc oxide from the melt [4]. The assembly comprises a metallic crucible which is surrounded by insulating media, e.g. zirconia, alumina etc. The iridium crucible is directly heated by an rf-coil, this means, that no external heater is present. The whole process is conducted in a high-pressure chamber to apply an oxygen-containing atmosphere. Elevated pressure is necessary in order to decrease the evaporation rate of zinc oxide. The dissociation is described by :

$$ZnO \leftrightarrow Zn + \tfrac{1}{2} O_2 \qquad (1)$$

Thermodynamical calculations have shown, that the total pressure of zinc oxide is $p_{tot} = 1.06$ bar at $T_f = 1975$ °C [5]. In addition, the evaporation of oxygen leads to remaining zinc that could easily react with iridium. Consequently, the overpressure is supplied by carbon dioxide, that decomposes at these high temperatures and results in a variable partial pressure of oxygen. In contrast to a pure oxygen atmosphere, when only one specific partial pressure of oxygen is present, the dissociation of carbon dioxide is also dependent on temperature and may lead to an "self-adjusting" oxygen partial pressure by choosing an appropriate composition of the atmosphere, e.g. by mixing carbon dioxide with argon [1]. Using a carbon dioxide atmosphere the iridium crucible remains stable even in an oxidizing atmosphere and the decomposition rate of zinc oxide is simultaneously reduced. This is one key parameter for melt crystal growth of zinc oxide in an iridium crucible.

Since the starting material offers a rather low powder density it has to be compressed for crystal growth in a crucible. This is also required because pre-melting is no option in presence of a seed crystal. The zinc oxide powder is first subjected to cold isostatic pressing in a mould at a pressure up to 2000 bar. To increase the density of the green compact it is subsequently heated at ca. 1100 °C for one day. Finally, a density of the compact of about 95 % of the theoretical value of zinc oxide can be obtained. Both polar directions have been already used for the seeds and other direction will be tested in future. The growth process can be divided into three basic steps:
1) melting and homogenization of the starting material,
2) adjusting the axial temperature gradient and partial melting of the seed,
3) crystallization.
During the first step heating proceeds up to a temperature above the melting point. Since temperature measurements are difficult at around 2000 °C, the event of melting is indirectly detected by a sudden change in temperature. The second step is established by changing the position of the crucible with respect to the rf-coil. This is accompanied by a constant temperature at the control point leading to a partial melting of the seed. After a second homogenization period the crystallization is initiated by a closed loop control of the temperature.

Currently single crystals of zinc oxide of 33 mm in diameter and up to 50 mm in length can be grown by the Bridgman technique. The limitation in diameter is only due to the use of a "standard" size. The crystalline properties were measured by X-ray diffraction and are represented by the rocking curve.

One critical issue in growth from the melt with the help of a crucible is contamination of the growing material by the crucible itself. Since purification of iridium is either difficult or expensive, we tried multiple usage of the crucible instead of single-use. This is complicated by the fact that the cylindrical part of the crucible has to separated from the cone in order to reveal the single crystal. After removal of the crystal the crucible is welded again and can be used for the second growth run. The impurity content has been investigated by secondary ion mass spectrometry (SIMS).

**DISCUSSION**

Crystal growth from the melt exhibits a number of features, which makes it superior to other growth techniques as far as single crystals are concerned. First, the ratio in density between the nutrient (e.g. vapor phase, solution) and the solid is closer to one when crystallizing from a melt compared to e.g. hydrothermal growth. This leads to a comparingly high crystallization rate. As a result the crystal volume produced per time makes melt growth the most economic process.

Large single crystals of zinc oxide are known to originate from growth methods from the vapor, hydrothermal growth as well as melt growth. Whereas the first two methods have been extensively studied for more than four decades [6], the first successful report on melt growth was released in 1999 [7]. To date actually only two groups worldwide are dealing with melt growth of zinc oxide.

## Hydrothermal method

Comparing the transport velocity of the above mentioned methods the lowest values are reported for hydrothermally grown crystals [8]. In alkaline aqueous solutions the crystal is free to grow in any direction and therefore, the growth rate of different faces has to be considered. Dependent on the choice of solvents as well as their concentration ratios the growth rate is normally below 0.2 mm/day for the [0001] direction [6, 9]. This low growth rate might be compensated by simultaneous growth of many crystals. Faster growth is observed in the directions perpendicular to the c-axis leading to typically platelet-like crystals. This strongly anisotropic growth may lead to growth sectors of different impurity content or defect density. Large single crystals up to 3 inch in diameter with excellent structural quality have been already demonstrated.

## Growth from the vapor

Physical vapor transport has been recognized as providing very low transport rates already in the 1970s [10]. Meanwhile numerous transport agents (e.g. $H_2$, C, $Cl_2$) have been tested and shown to not only increase the transport rate but also to result in high quality crystals [11, 12, 13]. Usually crystals grow in ampoules made of silica under wall contact. Crystals grown without wall contact show a needle-like habit with fastest growth along the c-axis. The growth velocity may reach up to a few hundred microns per hour. Single crystals up to two inch in diameter and 1 cm in thickness were reported [14].

## Melt growth

Melting of zinc oxide is still a sophisticated task due to several obstructive properties of the molten material. The relatively high total pressure (see Eq. 1) leads to a strong evaporation and requires an overpressure. Furthermore, there are only a few materials that can withstand more than 2000 °C in an oxidizing environment. To avoid the use of a crucible the "pressurized melt growth" has been described [15]. A skull melting equipment is used under elevated pressure up to 100 bar. This technique is based on induction heating, where eddy currents directly flow inside the material. The starting material is placed inside water-cooled copper fingers, preventing the rim from being melted. Finally, molten zinc oxide is contained in solid zinc oxide, that is also denoted as the "skull". Owing to the absent contact to a foreign material, except for the atmosphere, the purity should be as high as that of the starting material. Critical issues of this methods are temperature control and temperature gradients. Electromagnetic forces inside the melt influence strongly the flow pattern of the liquid. It is difficult to establish directional solidification, either by lowering the crucible or by using a pulling arrangement. Crystals as large as 2 inch in diameter have been prepared.

Several companies offer zinc oxide substrates based on the hydrothermal method and skull melting. By contrast, the Bridgman technique developed by the authors is still on a level of academic research. Single crystals up to 33 mm in diameter have been demonstrated with good crystalline perfection [4]. The full width at half-maximum (FWHM) of the X-ray rocking curve can be lower than 50 arcsec. However, a more realistic upper limit of the FWHM value is 300 arcsec, which is mainly due to low angle grain boundaries (Fig. 1). There is no indication of strain due to contact to the iridium crucible.

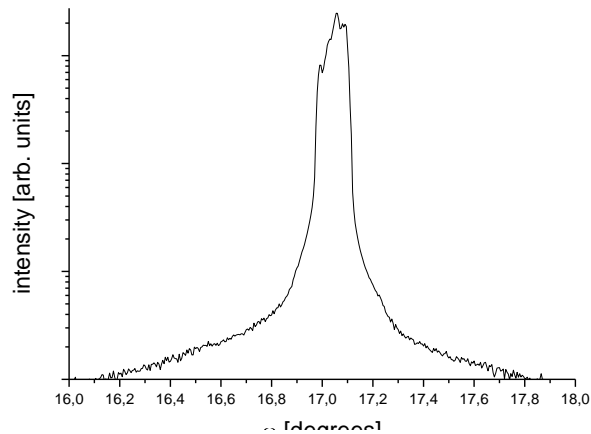

Figure 1: X-ray rocking curve for {0002} reflection, spot size 10 x 2 mm$^2$, open detector, log-scaling

During the growth process the crucible is nearly closed and there is no direct observation possible. So far the solid-liquid interface could not be detected in the grown crystals by different diagnostic techniques. The growth rate is estimated to be in the order of a few millimeters per hour, being one hundred times larger as for hydrothermal growth.

The purity of the grown crystal depends on the purity of the starting material and on the purity of the crucible. In order to identify the main sources for impurities three different crystals have been investigated by SIMS. In Table 1 a considerable improvement for most elements can be found when changing the starting material from 4N to 5N. Although to a lower extent this can also be seen when using a crucible for a second growth run. For this crystal two data are shown, from the seed end and the tail end respectively.

Table 1. Impurity analysis of ZnO crystals by SIMS (4N and 5N: purity of starting material, crucible re-use: 5N starting material and crucible used second time (upper row: seed end, lower row: tail end), HT: hydrothermal sample)

|  | Li | Na | Mg | Al | K | Ca | Cu | Ga | Si |
|---|---|---|---|---|---|---|---|---|---|
| ZnO 4N | 1E18 | 2E19 | 3E18 | 2E19 | 3E18 | 3E18 | 1E19 | 1E18 | 5E19 |
| ZnO 5N | 8E15 | 3E17 | 1E17 | 1E18 | 2E17 | 4E17 | 7E17 | 5E17 | 2E18 |
| ZnO crucible re-use | 4E15 | 4E17 | 1E17 | 5.1E17 | 2.3E17 | 1E17 | <5E17 | <1E16 | 1.7E18 |
|  | 8.3E15 | 1.2E17 | 3E16 | 3.2E17 | 4.2E16 | 2.6E16 | <5E17 | <1E16 | 3.4E17 |
|  |  |  |  |  |  |  |  |  |  |
| HT | 3E17 | 3E17 | 2E17 | 5E18 | 2E17 | 1E17 |  | 6E18 | 5E18 |

Most of the elements, e.g. Li, Na, Mg, K, Si, do not change in concentration by using the crucible for the second time. This means, that the crucible is probably not contaminated by these impurities. On the other hand, there is a difference for Al and Ga, the latter being below the detection limit. Since these elements act as donors in ZnO, their concentration should be well defined in terms of doping. For almost all impurities segregation is observed, leading to a lower concentration at the tail end compared to the seed end. For comparison a commercially purchased sample grown hydrothermally is added. Except for Li, Al and Ga one can find similar levels of contamination. Since for the hydrothermal method e.g. LiOH is used as a mineralizer, the concentration of Li is much higher in the crystals than in the melt grown ones. The high amount of donor-like impurities is in good agreement with temperature-dependent Hall effect measurements. Even though the electron concentration at room temperature for hydrothermal crystals is usually lower than that for Bridgman grown crystals, the measurements show for hydrothermal crystals a high degree of compensation.

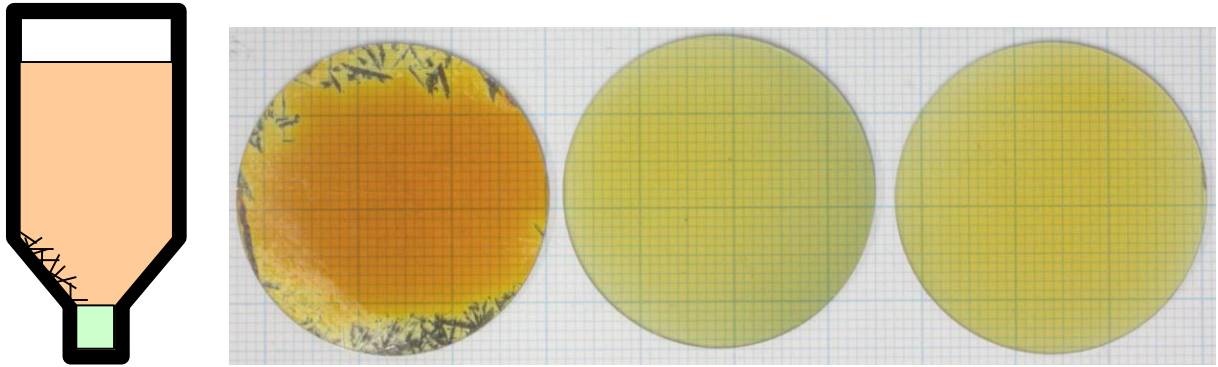

Figure 2. left: Schematic of the iridium crucible including seed and crystal, with iridium precipitation in the cone; right: three as-grown wafers from the bottom (left), middle (center) and the tail (right) part of the crystals

The analysis of impurities did not show the occurrence of iridium in the Bridgman grown crystal. Though iridium is even visible to the naked eye in some regions of the grown boule, it was not detected by SIMS in the investigated samples. Iridium can be found in the lower part, this means first grown part, of the crystal (Fig. 2). Probably the zinc oxide melt slowly dissolves the crucible, but the precipitation only occurs in the lower part of the setup. Therefore, it can be assumed, that the rate of which iridium is solved in the melt, decreases with time. Otherwise, it had to be found along the whole boule. Since the lower part, mainly the cone, is normally not used for wafer production, the iridium precipitation has only a low impact on the yield.

## CONCLUSIONS

Single crystal growth of zinc oxide is still under investigation, since commercially available substrates need further improvement and may play an essential role towards the development of p-type material. The Bridgman method is compared with the hydrothermal method, skull melting and growth from the vapor phase. The developed process offers several advantages e.g. up-scaling or dopant control. The growth of 2 inch crystals is currently under investigation and improvements of the crystal quality can be expected by adjustment of the thermal conditions during growth.

## ACKNOWLEDGMENTS

The authors gratefully acknowledge the contribution by S. Lautenschläger from Justus Liebig University Giessen for the SIMS measurements and would like to thank A. Kwasniewski for doing XRD.